\documentclass[aps,prl,twoside,twocolumn,notitlepage,nofootinbib,superscriptaddress,10pt]{revtex4-1}
\usepackage[CJKbookmarks, pdftex, bookmarksnumbered, bookmarksopen, colorlinks, citecolor=blue, linkcolor=blue]{hyperref}

\usepackage[english]{babel}
\usepackage{amstext,amssymb,amsfonts,bbm}
\usepackage{amsmath}
\usepackage[latin1]{inputenc}
\usepackage{graphicx}
\usepackage{epstopdf}
\usepackage{amssymb}
\usepackage{bbold}
\usepackage{amsthm}
\usepackage{tocvsec2}
\usepackage{enumerate}
\usepackage{subfigure}

\usepackage{bm}
\usepackage{color}
\usepackage{stmaryrd}

\newcommand{\Ignore}[1]{}

\newcommand{\be}{\begin{equation}}
\newcommand{\ee}{\end{equation}}
\newcommand{\bes}{\begin{eqnarray}}
\newcommand{\ees}{\end{eqnarray}}
\newcommand{\Ket}[1]{\left\vert #1\right\rangle}
\newcommand{\Bra}[1]{\left\langle #1\right\vert}

\newcommand{\MV}[1]{\left\langle #1 \right\rangle}

\begin{document}

\title{On the fate of the Hoop Conjecture in quantum gravity}

\author{Fabio Anz\`{a}$^{1}$, Goffredo Chirco$^{2}$}

\affiliation{$^1$Clarendon Laboratory, University of Oxford, Parks Road, Oxford OX1 3PU, United Kingdom.\\
$^{2,3}$Max Planck Institute for Gravitational Physics, Albert Einstein Institute, Am M\"{u}hlenberg 1, 14476, Potsdam, Germany.}

\date{\today}
\begin{abstract}
We consider a closed region $R$ of 3d quantum space modeled by $SU(2)$ spin-networks. Using the concentration of measure phenomenon we prove that, whenever the ratio between the boundary $\partial R$ and the bulk edges of the graph overcomes a finite threshold, the state of the boundary is always thermal, with an entropy proportional to its area. The emergence of a thermal state of the boundary can be traced back to a large amount of entanglement between boundary and bulk degrees of freedom. Using the dual geometric interpretation provided by loop quantum gravity, we interprete such phenomenon as a pre-geometric analogue of Thorne's ``Hoop conjecture'', at the core of the formation of a horizon in General Relativity.
\end{abstract}

\maketitle

\section{Introduction}

In statistical mechanics, small systems weakly coupled to a large bath are described by canonical ensembles when the composite system (system + bath) is in a microcanonical state\cite{StatMech1,StatMech2,StatMech3}. When we deal with a closed many-body quantum systems, the reduced density matrix of a small part of the system can be proven to be almost canonical, even if the state of the overall system is pure \cite{Typ1}. Such Gibbs-like behaviour emerges locally, in closed systems, as a direct consequence of the concentration of measure phenomenon. Its application to quantum statistical mechanics goes under the name of ``Canonical typicality''\cite{Typ2,Typ3,Typ4,Typ5,Typ6,Typ7}. It is a purely kinematic analysis on the Hilbert space of the system and it can be proven in full generality by means of Levy's Lemma\cite{led}. A brief summary of canonical typicality and of Levy's lemma can be found in Appendix \ref{app:typ} and \ref{app:levy}.\\

In a recent paper\cite{noi}, the authors showed that typicality arguments can be fruitfully applied to the Hilbert space of the spin-network states. In this context, typicality provides a remarkable tool to study the local behaviour of a quantum geometry, in a fully kinematic approach, hence consistently with the general covariant nature of the spin-network description.

In this letter we propose a radical shift of setting where the role of ``system'' and ``bath'' is played by the boundary and bulk degrees of freedom of a generic \emph{ball} of 3D quantum space. We prove that, whenever the Hilbert space of the boundary is much smaller than the bulk space, the reduced state of the boundary is always a thermal state, regardless what is the specific pure state of the whole region. In a series of recent works, such thermal states of the boundary have been interpreted as the pre-geometric equivalent of black hole configurations\cite{Etera0,eter2,eter3}.
Our general argument then comes in strong support to this vision, suggesting an information-theoretical origin for such proto-black-hole states.\\

There is more. We find that the typical character of such thermal state of the boundary is regulated by a \emph{threshold condition} which reproduces, at the pre-geometric level, the famous threshold mechanism of Thorne's Hoop Conjecture \cite{Hoop1,Hoop2}. We are then led to read this behaviour as an indirect proof of the Hoop Conjecture, in a purely quantum information-theoretic fashion.

\section{A 3-ball of quantum space}\label{sec:Qspace}

In many background-independent approaches to quantum gravity, Loop Quantum Gravity\cite{LQG1,LQG2,LQG3}, Spinfoams\cite{SF}, Group Field Theory\cite{GFT}, the microscopic quantum structure of space-time is described by purely combinatorial and algebraic variables, in terms of non-geometric structures realised by spin-networks\cite{spinnet1,spinnet2,spinnet3,spinnet4,spinnet5,spinnet6}. They are generic graphs $\Gamma$, made by vertices (or nodes) $v$ which are connected by edges $e$.  The edges are coloured with irreducible representations of the local gauge group of the theory. In this case the Lorentz group, gauge-fixed to SU(2). Therefore to each $e \in \Gamma$ one associates an $SU(2)$ irreducible representation (irrep) labelled by a half-integer $j_e \in \mathbb{N}/2$ called spin. The representation (Hilbert) space is denoted $V_{j_e}$ and has dimension $d_{j_e} = 2 j_e +1$. To each node of the graph $v$ one attaches an intertwiner $i_v$. This is an $SU(2)$-invariant map between the representation spaces $V_{j_e}$ associated to all the edges $e$ meeting at the vertex $v$,
\begin{align}
i_v: \,\, \otimes_{e\in v} V_{j_e}\to V_0 
\end{align}
In other words, $i_v$ is an invariant tensor, or a singlet state, between the representations attached to all the edges linked to the considered vertex. Once the $j_e$'s  are fixed, the intertwiners at the vertex $v$ actually form a Hilbert space $\mathcal{H}_v\equiv  \text{Int}_v [ \bigotimes_{e}V_{j_e}]$. A spin network state $|\Gamma, \{j_e\},\{i_v\}\rangle$ is defined then  as the assignment of representation labels $\{j_e\}$ to each edge and the choice of a vector $|\{i_v\}\rangle \in \otimes_v\mathcal{H}_v$ for the vertices. 

Therefore, for a given graph $\Gamma$, upon choosing a basis of intertwiners for every assignment of representations ${j_e}$, the spin networks provide a basis for the space of square-integrable wave-functional,  endowed with the natural scalar product induced by the Haar measure on $SU(2)$ 
\begin{align}
\mathcal{H}_{\Gamma} = L_2[SU(2)^E/SU(2)^V]= \bigoplus_{\{j_e\}} \bigotimes_v \mathcal{H}_v .
\end{align} 


%

\begin{figure}[h]
\includegraphics[width=3.6 in]{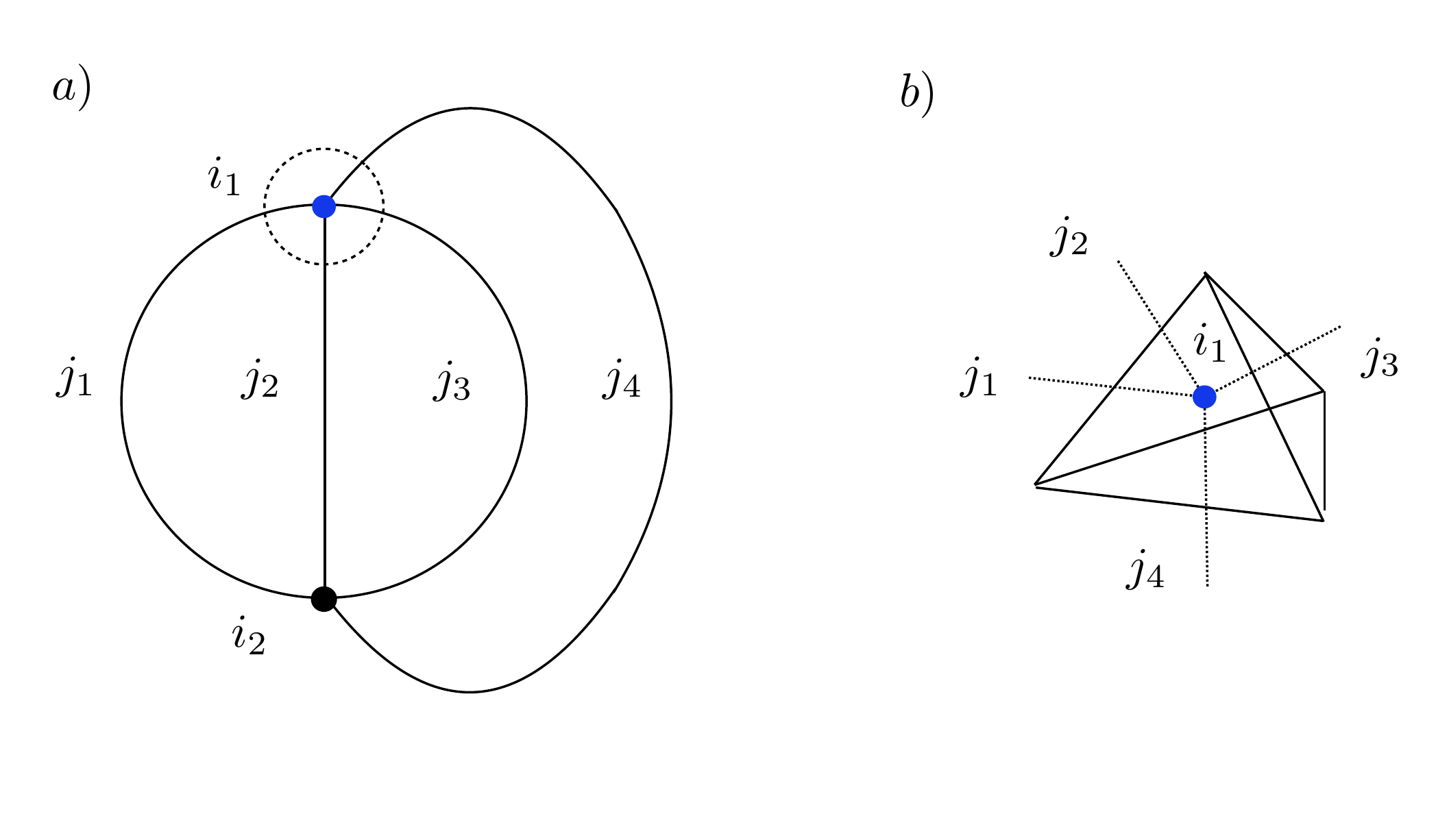}
\caption{A generic 4-valent closed spin network graph. b) Geometric interpretation of a single node as chunk of space. The links are dual to patches of surface whose area is proportional to the spin $j_i$. A node is dual to a chunk of space, with the shape of a polyhedron with a number of faces equal to the number of links around the node.}\label{fig:poly}
\end{figure}

From a geometrical point of view, given a cellular decomposition of a three-dimensional manifold, a spin-network graph with a node in each cell and edges connecting nodes in neighbouring cells is said to be dual to this cellular decomposition. Therefore, each edge of the graph is dual to a surface patch intersecting the edge and the area of such patch is proportional to the representation $j_e$. Analogously, vertices of a spin network can be dually thought of as chunks of volume (see \cite{poly} for the geometric interpretation of spin-networks states as collection of polyhedra). See Figure (\ref{fig:poly}) for an example.

In particular, a bounded region $R$, dual to a subregion of a generic spin network graph $\Gamma$, includes a finite set of vertices ($V_R$) and the edges connecting them ($E_R$). We define a boundary $\partial R$ as the set of edges which have only one end vertex laying in $R$. Their number is called $E$. Consistently, bulk edges are paths connecting vertices in $R$. We can picture $R$ as a 3-ball and $\partial R$ as its boundary 2-sphere punctured by the boundary edges.

\begin{figure}[h!]
\includegraphics[width=2.6 in]{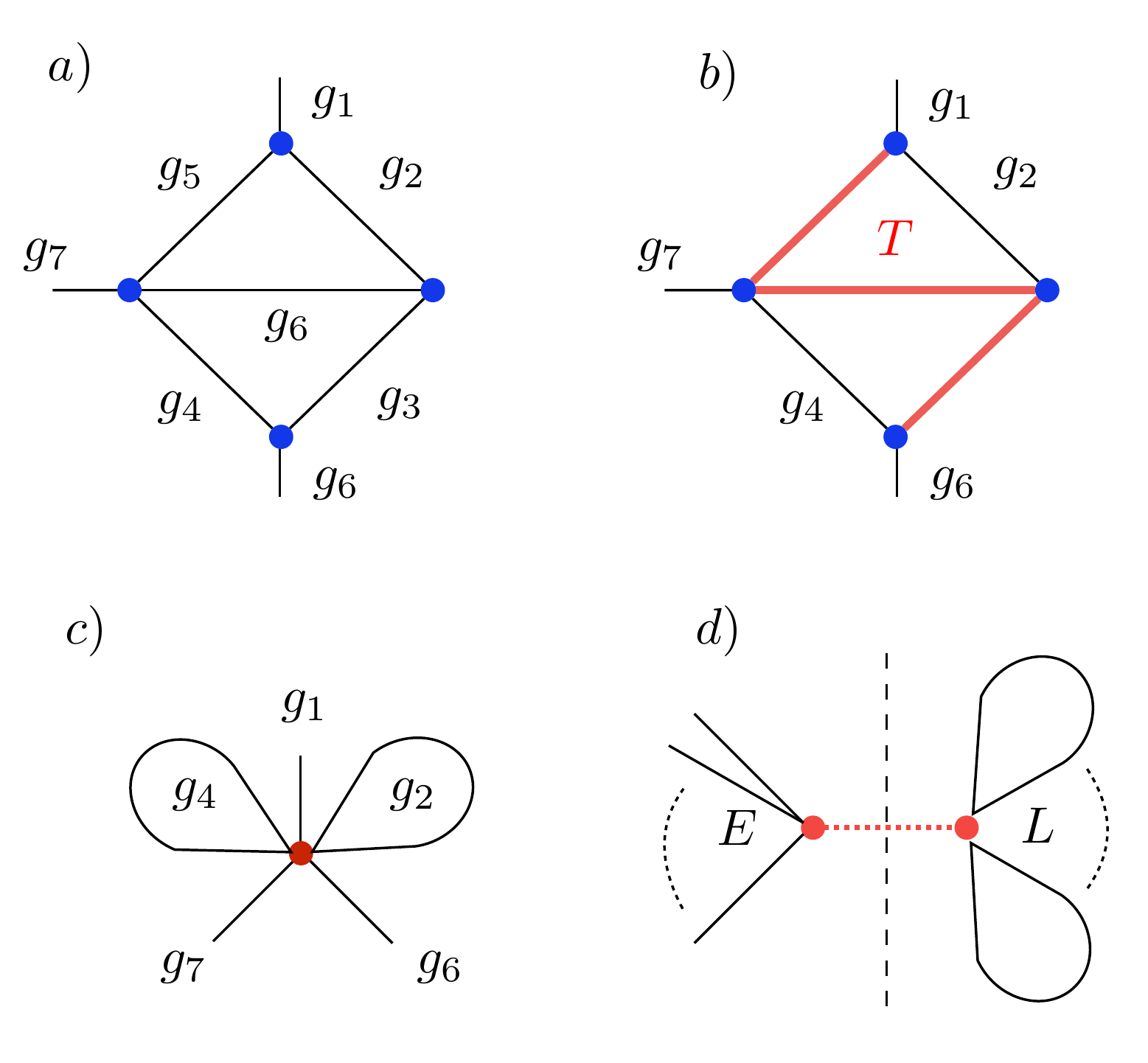}
\caption{a) Simple example of open graph reduced to a single intertwiner with two petals and three edges (c). b) One (out of six different) way to chose the set of holonomies comprising the maximal tree $T$. d) Graphic representation of the unfolding of the single vertex graph, with virtual link in red.}
\label{flowers}
\end{figure}

Thanks to the gauge invariance at each node inside $R$, we can gauge-fix this space and simplify the structure of bulk, without loosing any information \cite{eter1,eter2,eter3,eter4}. Remarkably,  the gauge-invariant Hilbert space associated the original graph $\Gamma_R$ is isomorphic to the gauge-invariant space defined on a different graph $\mathcal{F}_R$. Such graph consists of a single vertex intertwining the external edges of $\Gamma_R$ together with a certain number of loops $L$ which depends on the internal structure of $\Gamma_R$. The number of loops is $L=E_R-V_R +1$, where $V_{R}$ and $E_R$ are, respectively, the number of nodes and edges inside $\Gamma_R$. If $L=0$ there are no loops and $\Gamma_R$ has trivial topology. Such graph is usually called \emph{Flower graph}\footnote{To be precise, a flower graph is usually one that has only loops (the petals of a flower). However, we will use this convenient nomenclature to indicate a more generic graph with a certain number of loops and of external legs (the stems of a flower).} \\
Such gauge reduction isomorphism does not produce any coarse graining, nevertheless the correspondence between the original graph and its flower is many-to-one, due to the fact that the procedure discards the combinatorial information about the internal subgraph. An example is given in Figure (\ref{flowers})--(a-c).\\

The Hilbert space of $\mathcal{F}_R$ provides us with the most synthetic description for a region of quantum space with non-trivial internal degrees of freedom. 

\section{Separating Boundary and Bulk}

In order to make our argument transparent we consider a simplifed set-up by working with graphs with fixed spin representations $j_0$. This means that the gauge-invariant Hilbert space of our idealised 3-ball region is given by

\begin{align}
&\mathcal{H}_{\mathcal{F}_R} = \mathrm{Inv}_{SU(2)} \left[ V_{j_0}^{\otimes (E+2L)}\right]
\end{align}

%

Within $\mathcal{F}_R$ we identify the \emph{boundary} and \emph{bulk} degrees of freedom of the 3-ball respectively with open edges and internal loops. 
The separation is motivated by the different geometric information they carry. Loops are due to the presence of topological defects of the graph and carry localised ``curvature excitations'' at the vertices \cite{eter2,eter3}. On the other hand, the irreps carried by the open edges are dual to patches of the boundary surface.

Starting from such separation of degrees of freedom, we can think of $\mathcal{H}_{\mathcal{F}_R}$ as a \emph{bipartite} quantum mechanical system, with boundary and bulk subspaces weakly coupled by the constraint of $SU(2)$ gauge invariance at the vertex. In these terms our constrained space $\mathcal{H}_{\mathcal{F}_R}$ can be seen as embedded in a tensor product space 
\begin{align}\label{space}
\mathcal{H}_{\mathcal{F}_R} \subseteq \mathcal{H}\equiv \mathcal{H}_{\partial R} \otimes  \mathcal{H}_R,
\end{align} 
where we define the (unconstrained) boundary and bulk Hilbert spaces, respectively, as

\begin{align}
&\mathcal{H}_{\partial R} \equiv V_{j_0}^{\otimes E} && \mathcal{H}_R \equiv V_{j_0}^{\otimes 2L} \,\,\, .
\end{align}

The coupling induced by the request of gauge invariance translates into correlations between boundary and bulk. \\

Within $\mathcal{H}_{\mathcal{F}_R}$, a consistent reorganisation of boundary and bulk degrees of freedom is realised by considering the unfolding of the single vertex flower graph in a graph with two vertices, recoupling edges and loops irreps, linked by a virtual edge. An example is given in Figure (\ref{flowers})--(d) and it corresponds to the following re-writing of the gauge-invariant space 
 
\begin{align} \label{unfolded2}
\mathcal{H}_{\mathcal{F}_R}=  \bigoplus_{k}\mathcal{H}^{(k)}_E \otimes \mathcal{H}^{(k)}_L
\end{align}

where $k$ runs over the irreps of the virtual link. Moreover, $\mathcal{H}^{(k)}_E = V_k^{(E)} \cdot \mathcal{D}_k^E$ and $\mathcal{H}^{(k)}_L = V_k^{(L)} \cdot \mathcal{D}_k^{2L}$ and $\mathcal{D}_k^E$,$\mathcal{D}_k^{2L}$ counts the degeneracies of the space $V_k$ in the boundary and in the bulk recoupling. The details of the way in which such decomposition is done can be found in the supplementary material, along with the information about the dimensions of $\mathcal{H}_E^{(k)}$ and $\mathcal{H}_L^{(k)}$. Here we only need to fix the notation. If we write a generic decomposition as $\otimes_n V_j = \bigoplus_{k} V_k {}^jF_k^{n} $, we call ${}^jd_k^{n} \equiv \mathrm{dim} \,\,\, {}^jF_k^{n} $. All the dimensions needed in this paper can be written using ${}^jd_k^{n}$. More details can be found in the supplementary material.

\section{Typicality of the boundary}

As far as an observer \emph{external} to the region $R$ is concerned, the geometry of the region $R$ is described by the information measured on the boundary of such region $\partial R$. Such information is given by the \emph{reduced} density matrix obtained by taking the partial trace of the whole state $ |\varphi_{\Gamma_R}\rangle \langle \varphi_{\Gamma_R}|$ over the non-trivial holonomies around the loops (bulk degrees of freedom) $\rho_{\partial R} \equiv \text{Tr}_{L}[|\varphi_{\Gamma_R}\rangle \langle \varphi_{\Gamma_R}|]$.\\

Starting from the description of a region of quantum space given before, we focus on the state of the boundary.  Using the typicality tools (summarised in Appendix \ref{app:typ}) we prove that {\it whenever the dimension of the bulk Hilbert space exceeds the dimension of the boundary space, the reduced boundary state is always extremely close to the canonical state on the boundary $\Omega_{\partial R}$, regardless what is the global state of the whole region}. 

The {\it canonical state} of the boundary $\Omega_{\partial R}$ is defined in the following way. First we need $\mathcal{I}_{{\mathcal{F}_R}}$, the microcanonical state over $\mathcal{H}_{{\mathcal{F}_R}}$:

\begin{align}
&\mathcal{I}_{{\mathcal{F}_R}} \equiv \frac{1}{d_{{\mathcal{F}_R}}}{\mathbb{1}_{{\mathcal{F}_R}}} = \frac{1}{d_{{\mathcal{F}_R}}} P_{{\mathcal{F}_R}} ,
\end{align}
where $P_{{\mathcal{F}_R}}$ projects the states defined on $\mathcal{H}_{\partial R} \otimes \mathcal{H}_R$ onto the $SU(2)$ gauge invariant subspace $\mathcal{H}_{{\mathcal{F}_R}}$, while $d_{\mathcal{F}_R} = {}^{j_0}d_0^{E+2L}$ is the trace of $P_{{\mathcal{F}_R}}$. $\Omega_{\partial R}$ is the partial trace of $\mathcal{I}_{\mathcal{F}_R}$ over the bulk degrees of freedom:

\begin{align}
&\Omega_{\partial R} \equiv \mathrm{Tr}_L \,\, \mathcal{I}_{\mathcal{F}_R}
\end{align}

The quantity which we are interested in is the trace-distance\cite{Geos, Nielsen} between a generic reduced state $\rho_{\partial R}$ and $\Omega_{\partial R}$: $D(\rho_{\partial R},\Omega_{\partial R})$. Using the prescription developed in \cite{Typ1,Typ2} and summarised in Appendix \ref{app:typ}, it is possible to show that its average over the global Hilbert space $\mathbb{E}\left[D(\rho_{\partial R},\Omega_{\partial R})\right]$ satisfies 

\begin{align} \label{eq:bound}
&0 \leq \mathbb{E}\left[D(\rho_{\partial R},\Omega_{\partial R})\right] \leq \frac{d_{\partial R}}{\sqrt{d_{\mathcal{F}_R}}} 
\end{align}

Moreover, it can be proven that the fraction of states which are $\epsilon > 0$ away from this average is incredibly small

\begin{align}
\frac{\mathrm{Vol} \left[  \phi \in \mathcal{H}_{\mathcal{F}_R} \, \vert \, D(\rho_{\partial R}(\phi),\Omega_{\partial R}) \geq \eta \right] }{\mathrm{Vol} \left[ \phi \in \mathcal{H}_{\mathcal{F}_R} \right] } \leq \eta' \label{result}
\end{align}

where 
\begin{align}
& \eta' = 4\,  \mathrm{Exp} \left[ - \frac{2}{9\pi^3} d_{\mathcal{F}_R} \varepsilon^2 \right] && \eta < \varepsilon + \frac{1}{2} \frac{d_{\partial R}}{\sqrt{d_{\mathcal{F}_R}}}, 
\end{align}

Therefore, whenever the right-hand side of Eq.(\ref{eq:bound}) is much smaller than one, it will be concretely impossible to distinguish the actual reduced state $\rho_{\partial R}$ from $\Omega_{\partial R}$. The goal is then to evaluate this bound and find the regime where the distance is effectively pushed to zero.

\subsection{Evaluation of the bound}

Building on the technical results derived in \cite{eter2,eter3}, we work in the regime $E,2L \gg 1$, which goes along with our statistical approach. Using the expression derived
in \cite{eter2,eter3} we obtain

\begin{align}
&\frac{d_{\partial R}}{\sqrt{d_{\mathcal{F}_R}}} \sim (2j_0+1)^{\frac{E}{2}-L} [(j_0(j_0+1))(E+2L)]^{3/4}
\end{align}

The details can be found in the supplementary material. This quantity has a leading exponential behaviour, both in the number of external edges and in the number of loops over which we trace. The exponent becomes negative 
as soon as $E < 2L$.  Such fast decay is present for any choice of $j_0$, even far from the ``semiclassical regime''  $j_0 \gg 1$. For example in the smallest case $j_0=\frac{1}{2}$ we have

\begin{align}
&\frac{d_E}{\sqrt{d_{R}}} \sim 2^{\frac{E}{2}-L} \left(\frac{3}{4}\right)^{3/4}[(E+2L)]^{3/4}
\end{align}

This is not exactly a threshold behaviour but it is a fast exponential decay, which becomes faster as we approach the semiclassical regime $j_0 \gg 1$. In that regime the exponential decay of $\mathbb{E} \left[ D(\rho_E(\varphi),\Omega_E)\right]$ to zero approaches precisely a threshold behaviour, regulated by the condition

\begin{align}\label{eq:threshold}
&E < 2L
\end{align}

Will will discuss the physical meaning of this condition in the final section. For the time being, we note that the left-hand side is proportional to the total area while the right-hand side is proportional to the curvature excitations carried by the internal loops. Intuitively speaking, since the trace of the loop holonomy is the curvature around a path, in the semiclassical limit an increase in the number of internal loops corresponds to an increase of the gravitational energy \emph{density} within the bounded $R$ region dual to $\Gamma_R$. Therefore the condition \eqref{eq:threshold} can be loosely interpreted as an inequality relating the area of a closed surface with the gravitational energy density inside. This suggests a connection with the inequality in the Hoop Conjecture.

\section{The typical boundary state}

We now focus on the explicit form of the typical state of the boundary.  Starting from the decomposition of the constrained space given in Eq.(\ref{unfolded2}), a convenient basis in either of the two subspaces is labeled by three numbers, respectively $|k,m,a_k\rangle$ and $|k,m,b_k\rangle$, with $a_k, b_k$ running over the degeneracy of the irrep $V_k$ at fixed value of $k$\cite{noi,eter2}.

A basis for the single intertwiner space is then written as
\begin{align} \label{basis}
| k, a_k , b_k \rangle = \sum_{m=-k}^k \frac{(-1)^{k-m}}{\sqrt{d_k}}  |k, -m, a_k \rangle_E\, \otimes  | k, m, b_k \rangle_L
\end{align}
with $d_k=(2k+1)$.\footnote{Notice that in the chosen basis in \eqref{basis}, the flower graph can be represented as a \emph{bivalent} intertwiner, where the dependence on the spins $\{j_E\}, \{j_l\}$ is hidden in the re-coupled spin label $k$.}

Each basis state can be represented as a tensor product state on three subspaces (bi-orthogonal Schmidt decomposition \cite{peres}),
\begin{align} \label{biortho}
|k, a_k , b_k \rangle \equiv  |{k} \rangle_{V_k^{(E)}\otimes V_k^{(L)}} \otimes |{a_k}\rangle_{\mathcal{D}_k^E} \otimes |{b_k}\rangle_{\mathcal{D}_k^L}
\end{align}
where $k$ runs over the global angular momentum of the boundary and of the bulk, which have to be equal;  $|a_k\rangle$ labels a basis vector of $\mathcal{D}_k^E$ and $|b_k\rangle$ labels a basis vector of $\mathcal{D}_k^{2L}$.

%
The microcanonical state on the intertwiner space is given by the normalised density matrix, which can be written in the basis $\left\{ \Ket{k,a_k,b_k}\right\}$
\begin{align} \label{ro}
\mathcal{I}_{\mathcal{F}_R}&= \frac{1}{d_{\mathcal{F}_R}}\sum_{\substack{k, a_k, b_k}} \Ket{k,a_k,b_k}\Bra{k,a_k,b_k}.
\end{align}

Thanks to the specific basis chosen, the computation of the canonical state is straightforward. The partial trace of $\mathcal{I}_{\mathcal{F}_R}$ is easily computed:

\begin{align}
&\Omega_{\partial R}= \mathrm{Tr}_L \left[\mathcal{I}_{\mathcal{F}_R}\right] =  \sum_{\substack{k}} \frac{\,^{j_0}d_{k}^{(2L)}}{d_{\mathcal{F}_R}(2k+1)}\,\,\,{\mathbb{1}}_{{V}_k^E} \otimes \mathbb{1}_{\mathcal{D}_k^E}
\end{align}
where ${\mathbb{1}}_{{V}_k^E}$ and $\mathbb{1}_{\mathcal{D}_k^E}$ are, respectively, the identity over $V_k^E$ and $\mathcal{D}_k^E$.\\

\subsection{Behaviour of the canonical coefficient}
Now we study the canonical coefficient $W_k^E \equiv \frac{\,^{j_0}d_{k}^{(2L)}}{d_{\mathcal{R}}(2k+1)}$ in order to understand what is the predicted behaviour in the thermodynamic regime, which is $2L \gg E \gg 1$. Using the expression given in \cite{eter2,eter3}  we obtain

\begin{align}
&W_k^E \stackrel{2L \gg k}{\sim} (2j_0+1)^{-E} \left(1+\frac{E}{2L}\right)^{3/2} \left(\frac{k+1}{2k+1}\right) \label{eq:expansion}
\end{align}

It is interesting to see that $W_k^E$ depends on $k$ only through $\left(\frac{k+1}{2k+1}\right) \in [\frac{1}{2},1]$. This is a very mild dependence and as $k$ increases it fades away.

Eq.(\ref{eq:expansion}) holds in the regime $2L \gg k$. However, $k\in [0,k_{max}]$ where $k_{max} = j_0 \min (E,2L) = j_0E$. Therefore there are two cases: $j_0 \sim 1$ and $j_0 \gg 1$. In the first case $k_{max} = j_0 E \sim E \ll 2L$, hence \eqref{eq:expansion} holds for any possible $k$. In the second case $j_0 E$ might be larger than $2L$ and the expansion in Eq.(\ref{eq:expansion}) does not hold for all the $k$. In this last case, we check the asymptotic behaviour of $k_{max}$, which was studied in \cite{eter2,eter3}. It was shown that in $L\gg 1$, $k_{max} \sim \sqrt{2L}$ and ${}^{j_0}d_{k_{max}}^{2L} \sim \frac{(2j_0+1)^{2L}}{2L}$ and the proportionality coefficient depends on $j_0$. Therefore for $k\sim k_{max}$ we have

\begin{align}
&W_{k_{max}}^E \sim (2j_0+1)^{-E}[j_0(j_0+1)]^{3/2} \left( 1+ \frac{E}{2L}\right)^{3/2}
\end{align}

This confirms what we obtained before: the canonical coefficient $W_k^E$ depends in an extremely weak way on the topological defect $k$. Therefore $\Omega_{\partial R}$ is essentially a completely mixed state of the boundary. This picture can indeed be checked by computing the von Neumann entropy $S_{\mathrm{vN}}(\Omega_E) = - \mathrm{Tr} \Omega_E \log \Omega_E$: 

\begin{align}
&S_{\mathrm{vN}}(\Omega_E) \simeq E \log (2j_0+1) - \frac{3}{2} \log \left( 1+ \frac{ E}{2L}\right)
\end{align}

where $E \log (2j_0+1) = \log d_{\partial R}$ is the maximum allowed entropy. This confirms that $\Omega_{\partial R}$ is almost a microcanonical state, with an entropy proportional to its area $A = j_0 E$:

\begin{align}
&S(\Omega_{\partial R}) \simeq A \,\, \frac{\log{(2j_0+1)}}{j_0} - \frac{3}{2} \log\left(1+\frac{E}{2L}\right)
\end{align}

%
%
%
%
%
%
%
%

\section{Summary and discussion}

In this letter we study the properties of the boundary $\partial R$ of a region $R$ of 3D quantum space. We exploited the gauge invariant spin-network formalism to provide a synthetic description of $R$ in terms of a flower graph $\mathcal{F}_{R}$. In this picture, the boundary degrees of freedom are living on the external edges $E$ while the bulk degrees of freedom live on the internal loops $L$. By exploiting the so-called ``typicality approach'' we proved that, regardless what is the state of the whole system, if the number of internal loops exceed a certain threshold $2L > E$ the state of the boundary is always a microcanonical thermal state and its entropy is proportional to the area.

Despite its simplicity, the proposed model has an intriguing physical interpretation. 
The density matrix $\rho_{\partial R}$ of the boundary describes the state seen by a generic observer sitting outside the region. Such reduced state is mixed and it is not gauge invariant, as tracing over the loops holonomies necessarily breaks the closure constraint. The resulting \emph{closure defect} encodes the non trivial topological defect carried by the loops in $\Gamma_R$. 

The presence of a closure defect implies that there is no convex \emph{piecewise flat} polyhedron dual to the coarse intertwiner at the vertex. A dual convex polyhedron may exist only if embedded in a (homogeneous) \emph{curved space}, the curvature radius depending on the value of the closure defect \cite{eter1,eter2,eter3,eter4}.

As far as our virtual external observer is concerned then, a generic quantum region of space with non-trivial internal structure is defined by a mixed quantum state and described as a {\it curved} surface. 

Interestingly, the relation between the value of the closure defect and the curvature radius of the convex polyhedron dual to the vertex can be interpreted as a measure of the gravitational energy localised within the subgraph structure \cite{eter2,eter3}. Intuitively, an increase in the number of internal loops corresponds to an increase of the gravitational energy \emph{density} inside $R$. This would then correspond to an increase of the boundary surface curvature.\\

At the classical level, General Relativity predicts the increase of the gravitational energy \emph{density} of a region of space to be bounded by a threshold mechanism, responsible for the gravitational collapse which leads to the formation of a black hole. 
The black hole collapse is a universal phenomenon: it is scale invariant and valid for all masses, due to the equivalence principle. However, this behaviour is expected to be spoiled by quantum effects, as a black hole's mass smaller than its own Compton length would not exhibit the black hole hallmark, the event horizon.\\

The present  work shows that something similar happens at a pre-geometric level. When the number of internal loops exceed a certain threshold, the amount of information that we can retrieve from the boundary state vanishes. The only thing we can read-off is the sum of the spins, which is the total area of the region.

This vision is further supported by the explicit form of the threshold condition, $2L > E$, which reproduces the inequality at the root of Thorne's famous ``Hoop conjecture'' : {\it an horizon will form if and only if a mass $M$ gets compressed into a region with circumference in any direction $\mathcal{C}$ proportional to its mass}

\begin{align}
&\mathcal{C} \leq \mathcal{C}_{\mathrm{HC}} && \mathcal{C}_{\mathrm{HC}} \sim M
\end{align}
in units $G=c=1$.

In our information theoretic setting, when the information stored in the internal region is too much with respect to a limited set of boundary edges, this information can not be transmitted outside the region. This suggests an interpretation of this mechanism as an {\it information-theoretic collapse}, which does not rely on any causal geometric structure, but only on the \emph{entanglement} induced by the gauge invariance on the Hilbert space of the graph. 

Due to its extreme generality, such statistical relation between boundary and bulk degrees of freedom may reveal interesting realisations in the tensor network analysis of the holographic  geometry/entanglement correspondence \cite{holo1, holo2,holo3}, as well as in general context of complex networks \cite{ginestra1, ginestra2}.


\section*{Acknowledgements}

The authors are grateful to Daniele Oriti, Mingyi Zhang and Pietro Don\`a for useful discussions. F.A. would like to thank the ``Angelo Della Riccia'' foundation for providing support for this research.\\

\appendix

\section{Typicality}  \label{app:typ}

In this appendix we give a brief summary of the result achieved in \cite{Typ1}. Suppose we have a generic \emph{closed} system, which we call ``universe'', and a bipartition into ``small system'' $S$ and ``large environment'' $E$. The universe is assumed to be in a pure state. We also assume that it is subject to a completely arbitrary \emph{global constraint} $\mathcal{R}$. For example, in the standard context of statistical mechanics it can be the fixed energy constraint. Such constraint is concretely imposed   by restricting the allowed states to the subspace $\mathcal{H}_{\mathcal{R}}$ of the states of the total Hilbert space $\mathcal{H}_U$ which satisfy the constraint $\mathcal{R}$:
\begin{align}
\mathcal{H}_{\mathcal{R}} \subseteq \mathcal{H}_{U} = \mathcal{H}_E \otimes \mathcal{H}_S \,\,\, .
\end{align}
$\mathcal{H}_S$ and $\mathcal{H}_E$ are the Hilbert spaces of the system and environment, with dimensions $d_S$ and $d_E$, respectively. We also need the definition of the canonical state of the system $\Omega_S$, obtained by tracing out the environment from the microcanonical (maximally mixed) state $\mathcal{I}_{\mathcal{R}}$

\begin{align}
&\Omega_S  \equiv \mathrm{Tr}_E [\mathcal{I}_{\mathcal{R}}] &&\mathcal{I}_{\mathcal{R}} \equiv \frac{\mathbb{1}_{\mathcal{R}} }{d_{\mathcal{R}}}
\end{align}

where $\mathbb{1}_{\mathcal{R}} $ is the projector on $\mathcal{H}_{\mathcal{R}}$, and $d_{\mathcal{R}} = \mathrm{dim} \,\mathcal{H}_{\mathcal{R}}$. This corresponds to assigning {\it a priori} equal probabilities to all states of the universe consistent with the constraints ${\mathcal{R}}$.

In this setting, given an arbitrary pure state of the universe satisfying the constraint $\mathcal{R}$, i.e. $\Ket{\phi} \in \mathcal{H}_{\mathcal{R}}$, the reduced state $\rho_S(\phi) \equiv \mathrm{Tr}_E[\Ket{\phi}\Bra{\phi}]$ will {almost} always be very close to the canonical state $\Omega_S$.

Concretely, such a behaviour can be stated as a theorem \cite{Typ1}, showing that for an arbitrary $\varepsilon > 0$, the distance between the reduced density matrix of the system $\rho_S(\phi)$ and the canonical state $\Omega_S$ is given probabilistically by

\begin{align}
\frac{\mathrm{Vol} \left[  \phi \in \mathcal{H}_{\mathcal{R}} \, \vert \, D(\rho_S(\phi),\Omega_S) \geq \eta \right] }{\mathrm{Vol} \left[ \phi \in \mathcal{H}_{\mathcal{R}} \right] } \leq \eta' \label{result}
\end{align}
where the {\it trace-distance} $D$ is a metric \footnote{We use the definition $D(\rho_1,\rho_2) = \frac{1}{2} \sqrt{(\rho_1 - \rho_2)^{\dagger}(\rho_1 - \rho_2)}$.} on the space of the density matrices \cite{Geos, Nielsen}, while
\begin{align}
& \eta' = 4\,  \mathrm{Exp} \left[ - \frac{2}{9\pi^3} d_{\mathcal{R}} \varepsilon^2 \right] && \eta = \varepsilon + \frac{1}{2} \sqrt{\frac{d_S}{d_E^{\mathrm{eff}}}}, 
\end{align}
with the effective dimension of the environment defined as
\begin{align}
 d_E^{\mathrm{eff}} \equiv \frac{1}{\mathrm{Tr}_E \left[ \left( \mathrm{Tr}_S \mathcal{I}_{\mathcal{R}} \right)^2\right] } \geq \frac{d_{\mathcal{R}}}{d_S} \label{deff}.
\end{align}
The bound in \eqref{result} states that the fraction of the volume of the states which are far away from the canonical state $\Omega_S$ more than $\eta$ decreases exponentially with the dimension of the ``allowed Hilbert space'' $d_{\mathcal{R}} = \mathrm{dim} \mathcal{H}_{\mathcal{R}}$ and with $\varepsilon^2 = \left(\eta - \frac{1}{2} \sqrt{\frac{d_S}{d_E^{\mathrm{eff}}}}\right)^2$. This means that, as the dimension of the Hilbert space $d_{\mathcal{R}}$ grows, a huge fraction of states gets concentrated around the canonical state. 
%

The proof of the result relies on the \emph{concentration of measure phenomenon}. The key tool to prove this result is the \emph{Levy} lemma, which we briefly report for completeness in Appendix \ref{app:levy}.\\

\section{Levy's lemma} \label{app:levy}
In order to better understand typicality it is useful to look at its most important step, which is the so-called Levy-lemma. Take an hypersphere in $d$ dimensions $S^{d}$, with surface area $V$. Any function $f$ of the point which does not vary too much 

\begin{align}\nonumber
&f :  S^d \ni \phi \to f(\phi) \in \mathbb{R} &&|\nabla f| \leq 1
\end{align}

will have the property that its value on a randomly chosen point $\phi$ will approximately be close to the mean value.
\begin{align}\nonumber
\frac{\mathrm{Vol}\left[ \phi \in S^d \, : \, f(\phi) - \MV{f} \geq \epsilon \right]}{\mathrm{Vol}\left[ \phi \in  S^d \right]} \leq 4 \, \mathrm{Exp} \left[ - \frac{d+1}{9 \pi^3} \epsilon^2 \right]
\end{align}

Where $\mathrm{Vol}\left[ \phi \in S^d \, : \, f(\phi) - \MV{f} \geq \epsilon \right]$ stands for ``the volume of states $\phi$ such that $f(\phi) - \MV{f} \geq \epsilon$''. $\MV{f}$ is the average of the function $f$ over the whole Hilbert space and $\mathrm{Vol}\left[ \phi \in  S^d \right]$ is the total volume of the Hilbert space. Integrals over the Hilbert space are performed using the unique unitarily invariant Haar measure.\\

The Levy lemma is essentially needed to conclude that all but an exponentially small fraction of all states are quite close to the canonical state. This is a very specific manifestation of a general phenomenon called ``concentration of measure'', which occurs in high-dimensional statistical spaces \cite{led}.

The effect of such result is that we can re-think about the ``a priori equal probability'' principle as an ``apparently equal probability'' stating that: as far as a small system is concerned almost every state of the universe seems similar to its average state, which is the maximally mixed state $\mathcal{E}_{\mathcal{R}} = \frac{1}{d_{\mathcal{R}}}\mathbb{1}_{\mathcal{R}}$.\\

\end{document}